\documentclass[12pt]{article}
\usepackage[utf8]{inputenc}
\usepackage{amsmath}
\usepackage{graphicx}
\usepackage{adjustbox}
\usepackage{float}
\usepackage{pdfpages}
\usepackage{setspace}
\usepackage{amssymb}
\usepackage{subfigure}
\usepackage{hyperref}
\usepackage{tikz}
\usepackage{caption}
\usepackage{bbm}
\usepackage{titlesec}
\usetikzlibrary{arrows.meta}
\usepackage[margin=1in]{geometry}
\usepackage{lscape}
\usepackage{pgfplots}
\usepackage{orcidlink}
\pgfplotsset{compat=1.8}
\usepgfplotslibrary{statistics}

\definecolor{mygray1}{gray}{0.9}
\definecolor{mygray2}{gray}{0.7}
\definecolor{mygray3}{gray}{0.3}
\definecolor{mygray4}{gray}{0.1}
\title{A Note on Uncertainty Quantification for Maximum Likelihood Parameters Estimated with Heuristic Based Optimization Algorithms}
\author{Zachary Porreca\orcidlink{0000-0002-2213-1437}\footnote{CLEAN Unit, Via Röntgen n. 1, Milan 20136 (Italy), zachary.porreca@unibocconi.it}\\ Bocconi University}

\date{January 2024}

\begin{document}

\maketitle

\vspace{-4mm}

\begin{abstract}
	
\singlespacing
\noindent 

Gradient-based solvers risk convergence to local optima, leading to incorrect researcher inference. Heuristic-based algorithms are able to ``break free" of these local optima to eventually converge to the true global optimum. However, given that they do not provide the gradient/Hessian needed to approximate the covariance matrix and that the significantly longer computational time they require for convergence likely precludes resampling procedures for inference, researchers often are unable to quantify uncertainty in the estimates they derive with these methods. This note presents a simple and relatively fast two-step procedure to estimate the covariance matrix for parameters estimated with these algorithms. This procedure relies on automatic differentiation, a computational means of calculating derivatives that is popular in machine learning applications. A brief empirical example demonstrates the advantages of this procedure relative to bootstrapping and shows the similarity in standard error estimates between this procedure and that which would normally accompany maximum likelihood estimation with a gradient-based algorithm. 
\end{abstract}


 
 \clearpage
\doublespacing
\vspace{-20mm}
\section{Introduction}
Much of modern applied economics relies on the estimation of parametric models with optimization algorithms. Maximum likelihood estimation is one such approach that relies upon algorithmically selecting parameters that maximize the conditional joint density of observed data (Greene 2018). With advances in computing, data sets are growing exponentially larger and models are being estimated on much broader parameter spaces. The recent widespread popularity of fixed effects estimation provides a timely example of this phenomenon. Many computational optimization platforms default to gradient-based algorithms to find model solutions. These solvers calculate the gradient (a vector of first derivatives for the objective function with respect to parameters) and use this to guide their search for the model's optimum\footnote{In short this is also true of algorithms that make use of Hessian matrices}. Gradient-based solvers face an increased risk of convergence to a local optima, rather than the true global optimum, in complex parameter spaces. Local optima convergence can in turn lead the researcher to incorrect model inference\footnote{Chen (2023) proposes an estimator for this same context that relies on the multiple evaluations from different initial starting values, so as to allow inference from the ``range" of local optima.}.

In dealing with this problem, researchers largely have only two tools at their disposal. For one, the researcher may choose to iterate through repetitions of the optimization procedure using different starting points, selecting the solution that yields the optimal likelihood across all of their iterations (as in Jourdain et al. (2021)). Alternatively, researchers often employ non-gradient based heuristic solvers that employ elements of stochastic search to find the true global optimum. Examples include Simulated Annealing (see Goffe et al. (1994) for a discussion of that algorithm's use in economics) and Particle Swarm (see Khalaf \& Lin (2021) for a discussion of that algorithm's applied use in economics). These global search algorithms are able to avoid getting trapped in local optima, searching the broader parameter space to find the optimum point. Gilli \& Winker (2009) provide a nuanced introduction to the usage of these types of algorithms and their applications to economics.

These algorithms are able to find global optima (often regardless of the initial starting point\footnote{Another sub-optimal feature of gradient-based solvers is their dependence on starting values. Different initial points can result in different parameter estimates.}), but do this at the expense of computational time, as noted in Goffe et al. (1994). This is a particularly troubling trade-off, as resampling methods can be a valuable asset in approximating covariance matrices. In the context of maximum likelihood estimation, Wooldridge (2010) provides convenient functional forms for approximations of the covariance matrix from the information utilized by gradient based solvers. The asymptotic covariance matrix for maximum likelihood derived parameters, ~$\hat{\theta}$, can be approximated as:

\begin{equation}
    \hat{Avar}(\hat{\theta})\approx\bigg[-\bold{H}(\hat{\theta})\bigg]^{-1}\approx\bigg[\bold{g}(\hat{\theta})\bold{g}(\hat{\theta})^T\bigg]^{-1}
\end{equation}

The covariance matrix can be approximated as either the inverse of the negative Hessian matrix evaluated at estimated optimal parameters or as the inverse of the outer product of the gradient/score evaluated at estimated optimal parameters. Greene (2018) provides a corresponding robust estimator of the approximated asymptotic covariance matrix, that leverages a sandwich structure. 

\begin{equation}
    \hat{Avar}(\hat{\theta})\approx\bigg[-\bold{H}(\hat{\theta})\bigg]^{-1}\bigg[\bold{g}(\hat{\theta})\bold{g}(\hat{\theta})^T\bigg]\bigg[-\bold{H}(\hat{\theta})\bigg]^{-1}
\end{equation}

Heuristic solvers do not provide approximations of the gradient/score or Hessian matrix that these commonly used textbook approximations of the covariance matrix rely upon. Further, given the length of time taken for these sorts of algorithms to converge, common alternatives based on resampling methods (such as that demonstrated in Kalai \& Vempala (2006)) that rely on repetition of the estimation procedure may be practically infeasible. 

Here, I provide a simple two-step solution to this problem that makes use of techniques now common in machine learning applications. This procedure entails estimating the parameters with the heuristic solver and then inputting those estimated parameters into automatic differentiation software to calculate the gradient/Hessian at that estimated optimal point to be used in computing an approximation of the covariance matrix\footnote{This method is functionally similar to that employed in Amilon (2003). There, the author estimated parameter values using the heuristic based Simulated Annealing algorithm and then input those parameters as initial values for a gradient-based solver.}. This method is able to estimate the covariance matrix for differentiable functions that gradient-based algorithms could otherwise be applied to (at the risk of local optima convergence). Further, automatic differentiation is able to provide more precision and accuracy than the numerical differentiation methods that are prone to rounding errors and are often employed \emph{within} the solvers that are used for gradient-based optimization (Baydin et al. 2018).

To demonstrate this method, I replicate the model and estimates from a published applied economics paper with several different optimization algorithms. Focusing on the heuristic based Simulated Annealing algorithm, I then bootstrap standard errors for parameter estimates and compare these to the standard errors derived from applying an automatic differentiation method to evaluate the likelihood function at the estimated optimal parameters and evaluate the similarity between the two resulting vectors. Computational time required for the two-step procedure is exponentially less than that required for bootstrapping. The two-step procedure took under one hour for convergence of the simulated annealing algorithm (from a randomly generated starting point) and approximation of the standard errors, compared to nearly a week of computation time for a bootstrap procedure of only 150 repetitions. For expositional purposes, I chose to utilize a model and data that are quite simple. The gains in reduced computation time that can be achieved by using the proposed two-step procedure, making use of tools common in the machine learning space, have the potential to be genuinely impactful to the researcher in settings in which the risk of convergence to a local optima makes the use of a gradient-based solver infeasible.

\section{Automatic Differencing as a Solution}
Automatic differentiation is a method of computing the derivatives of a function by decomposing it into a series elementary operations of the programming language and combining the derivatives of those operations via the chain rule to calculate the overall derivative of the function (Verma 2000). This methodology allows for extremely precise estimation of numerical derivatives and has found much recent appeal in machine learning applications (Baydin et al. 2018). Baydin et al. (2018) provide an overview of this method, its differences from numerical and symbolic differentiation, and its applications.

This procedure can be applied to approximate the covariance matrix for parameters estimated with heuristic algorithms. The gradient/score of a likelihood function is simply a vector of its first derivatives evaluated at a given parameter vector, while the Hessian matrix is simply a matrix of the function's second derivatives estimated a given parameter vector (Wooldridge 2010). The gradient/Hessian asymptotically approach zero at the optimal point (Wooldridge 2010). Using this premise, typical gradient based algorithms search for their optimal parameter vector. In our context, the gradient/Hessian evaluated at that optimal parameter vector is used in approximating the covariance matrix. 

The two-step procedure proposed here entails finding that same parameter vector with the use of a heuristic based algorithm and then simply evaluating the same objective/likelihood function at that estimated parameter vector with automatic differencing to provide a (potentially more accurate) approximation of the gradient/Hessian. The calculated vector or matrix can then be used to approximate the covariance matrix with one of the textbook formulations mentioned above. In this next section, I demonstrate this strategy with the data and multinomial logit model from a published paper.  

\section{An Empirical Example}
To demonstrate the usage and relative advantage of this procedure, I replicate the multinomial logit model specification of Nguyen-Van et al. (2017). In that study, the authors model the choice of tea variety to cultivate for Vietnamese farmers on a vector of household characteristics (Nguyen-Van et al. 2017). For simplicity, I replicate their ``without unobserved farmer heterogeneity"  multinomial logit model with the same data set. This model entails the estimation of 75 parameters across 216 observations: a parameter space likely to contain multiple local optima. First, to ensure that the model was accurately represented for the ensuing exercise, using the ``mlogit" package in R (Croissant 2020) the parameter estimates and standard errors from the original paper were replicated successfully.\footnote{The ``mlogit" package employs the Newton-Raphson (a gradient-based algorithm) by default and computes standard errors as the inverse of the negative Hessian matrix.} Similarly, the model and results were replicated in both R and Julia (the language in which automatic differencing for calculating the gradient/Hessian is conducted in) by optimizing manually coded representations of that paper's underlying model\footnote{Both of these replications employed the gradient-based BFGS algorithm.}. 

Proceeding with the actual empirical bench-marking exercise, I first estimate their model using the Simulated Annealing algorithm with a vector of random values as a starting point for the algorithm\footnote{These random values were drawn from a normal distribution with a mean of 0 and a standard deviation of 1.} to benchmark the computational time necessitated by the two-step procedure proposed here\footnote{Optimization with the Simulated Annealing algorithm was conducted in R using the ``GenSa" package detailed in Xiang et al. (2013) with rather large bounds (-20 to 20) placed around parameter estimates to expedite the procedure.}  The two-step procedure took under one hour to complete both the algorithm's convergence and the automatic differentiation needed to compute standard errors\footnote{In total, the two-step procedure took 53.7 minutes for the Simulated Annealing algorithm to converge (using R) and less than 1 second to compute standard errors using the inverse of the outer product of the gradient representation of the covariance matrix (using the Zygote automatic differentiation platform in Julia).}. This is significantly less than the time necessitated by the bootstrapping procedure. Only iterating through 150 bootstrap repetitions required just over six days of computation time. 

The Euclidean distance between the vector of standard errors generated by the two-step procedure and the vector estimated through bootstrapping is ~$33.198$. Across 100,000 iterations of a permutation test (as discussed in Legendre \& Legendre 2012), not a single iteration produced a Euclidian distance greater than that observed between the vector of standard errors derived from bootstrapping and that derived from the two-step procedure. For comparison, when comparing the vector of standard errors from the two-step procedure to standard errors derived directly from the gradient computed and utilized in the computations of more typical gradient based algorithms, the Euclidian distance between the Simulated Annealing vector and a vector from a BFGS gradient is ~$19.337$ and the distance between the Simulated Annealing Vector and a vector from a Newtonian gradient is ~$17.547$. Permutation tests have only 1.72\% and 1.69\% of permuted samples providing lower distances than that between the Simulated Annealing vector and the BFGS and Newtonian vectors respectively. The standard errors derived from the two-step procedure are extremely similar to those that would be derived from traditional gradient-based optimization methods, while the standard errors derived from bootstrapping are quite different. The two-step procedure is able to estimate standard errors for parameters estimated with heuristic algorithms that are closer to those of traditional gradient-based estimates than bootstrapping is able to. 

\section{Conclusion}
This note is meant to demonstrate how applied researchers can quantify uncertainty in their maximum likelihood parameter estimates, despite needing to make those estimates with non-gradient based solvers. There is no requisite trade-off between finding a global optimum and being able to estimate standard errors. The procedure outlined here makes use of a computational method, common in computer science and machine learning, for calculating a function's derivatives that is faster and more precise that other computational methods and is less error prone than manually doing so. This two-step procedure is exponentially faster than resampling based inference for the same heuristic-based solver's estimates. The procedure demonstrated in this note allows for applied researchers to conduct proper inference on global optima in the context of maximum likelihood estimation.

\end{document}